\documentclass[conference,10pt]{IEEEtran}
\IEEEoverridecommandlockouts

\usepackage[colorlinks=true,       
            linkcolor=blue,        
            citecolor=blue,        
            urlcolor=blue,         
            bookmarks=true,        
            pdfborder={0 0 0}]{hyperref}

\usepackage[all]{hypcap} 

\usepackage{cite}
\usepackage{amsmath,amssymb,amsfonts}
\usepackage{algorithmic}
\usepackage{graphicx}
\usepackage{textcomp}
\usepackage{xcolor}
\usepackage{bm}
\usepackage{mathtools}
\usepackage{bbm}
\usepackage{balance}
\usepackage[caption=false,font=footnotesize]{subfig}

\newcommand{\vect}[1]{\bm{#1}}
\newcommand{\diff}{\,\mathrm{d}}
\newcommand{\Ind}{\mathbbm{1}}
\newcommand{\E}{\mathbb{E}}
\newcommand{\eqdef}{\coloneqq}

\graphicspath{{Figures/}}

\usepackage{tikz}
\usetikzlibrary{3d, decorations.pathmorphing, arrows.meta}

\usepackage{booktabs}

\begin{document}


\title{Spatiotemporal First-Arrival Modeling and Parameter Estimation in Drift-Diffusion Molecular Channels
\thanks{This work was supported in part by the National Science and Technology Council (NSTC), Taiwan, under Grant NSTC 113-2115-M-008-013-MY3.}
}

\author{
\IEEEauthorblockN{Yun-Feng Lo}
\IEEEauthorblockA{\textit{School of Electrical and Computer Engineering}\\
Georgia Institute of Technology\\
Atlanta, GA, USA\\
yun-feng.lo@gatech.edu}
\and
\IEEEauthorblockN{Yen-Chi Lee}
\IEEEauthorblockA{\textit{Department of Mathematics}\\
National Central University\\
Taoyuan, Taiwan\\
yclee@math.ncu.edu.tw}
}

\maketitle

\begin{abstract}
We derive a closed-form joint distribution of the first arrival time (FAT) and first arrival position (FAP) in drift-diffusion molecular communication (MC) channels. In contrast to prior studies that analyze FAT or FAP in isolation, our framework explicitly captures the spatiotemporal coupling inherent in multidimensional transport. Building on this derivation, we compute the Fisher information matrix (FIM) and demonstrate that estimation accuracy for diffusivity scales proportionally with the spatial dimension, enabling increased sensitivity in higher-dimensional environments. Furthermore, we show that lateral drift—which is unobservable from timing data alone—can be recovered via a closed-form Maximum Likelihood Estimator (MLE) with a simple physical interpretation. Leveraging this spatial degree of freedom, we propose Drift Shift Keying (DSK), proving that joint receivers can reliably detect signals that are undetectable to timing-only receivers due to identical marginal FAT distributions. These results highlight the significant potential of spatiotemporal processing for future nanoscale communication and sensing.
\end{abstract}


\section{Introduction}

Molecular communication (MC) via diffusion is a biologically inspired paradigm for nanoscale information exchange, wherein information carriers propagate via Brownian motion superimposed with background flow~\cite{farsad2016comprehensive,nakano2013molecular}. The \emph{first arrival time} (FAT) is a critical metric for timing-based MC systems, widely modeled by the inverse Gaussian (IG) distribution~\cite{chhikara2024inverse,kadloor2012molecular}. Extensive research has leveraged FAT statistics for synchronization and channel parameter estimation~\cite{lin2015maximum,hsu2014timing}.

In multidimensional environments, however, molecules arrive at spatially distributed locations on the receiver surface due to lateral diffusion. This necessitates the study of the \emph{first arrival position} (FAP), which complements temporal information. Recent work has characterized FAP statistics using stochastic partial differential equations~\cite{lee2024characterizing}, showing that spatial arrival profiles encode key geometric features such as drift direction.

Despite these advances, most existing literature treats FAT and FAP separately. In this work, we develop a \emph{joint} characterization of FAT and FAP under constant drift and isotropic diffusion. Our derived joint distribution captures the interaction between detection latency and spatial dispersion. While spatiotemporal designs have been explored in joint time--concentration modulation~\cite{mirkarimi2020capacity}, a rigorous characterization of the FAT--FAP coupling remains an open problem. To the best of our knowledge, this is the first closed-form joint FAT--FAP characterization under constant drift.

Beyond channel modeling, this work leverages joint statistics to bridge the gap between theoretical physics and system design. We derive the Fisher Information Matrix (FIM) and establish the Cramér-Rao Lower Bounds (CRLB) for channel parameters, revealing three key system-level insights:
\begin{itemize}
    \item \textbf{Observability:} We prove that lateral drift, which is mathematically unobservable to time-only receivers, can be efficiently estimated using joint statistics. We provide a closed-form Maximum Likelihood Estimator (MLE) with a simple physical interpretation: the ratio of total lateral displacement to cumulative arrival time.
    \item \textbf{Sensitivity:} We demonstrate that estimation accuracy for diffusivity scales linearly with the spatial dimension $D$, enabling high-precision sensing with fewer molecules compared to traditional timing-based methods.
    \item \textbf{Modulation:} We propose \emph{Drift Shift Keying (DSK)}, a novel modulation scheme that exploits the spatial degree of freedom. We show that while time-only receivers fail completely in this regime ($P_e=0.5$), joint receivers can achieve reliable communication, thereby unlocking new spatiotemporal diversity gains.
\end{itemize}

\section{System Model} \label{sec:system}

We consider a single-input single-output (SISO) MC system in a $D$-dimensional unbounded fluid medium characterized by constant drift and diffusion. A transmitter (TX) releases molecules that are subsequently absorbed by a receiver (RX). 
We adopt a dimensionless formulation by normalizing length by the TX--RX distance $\lambda$ and time by $\lambda/v_p$, where $v_p > 0$ denotes the drift component perpendicular to the receiver plane. 
Under this scaling, the perpendicular drift is normalized to $1$, and the drift vector is given by $\vect{v} = (1, v^{(2)}, \ldots, v^{(D)})$. 
Unless otherwise stated, all parameters in this section are dimensionless.

The TX and RX are modeled as parallel $(D-1)$-dimensional hyperplanes separated by a unit distance along the $x^{(1)}$ axis. 
The TX emits a single molecule at time $t=0$ from the location
\begin{equation}
    \vect{x}_0 = (0, x_0^{(2)}, \ldots, x_0^{(D)}),
\end{equation}
and the RX absorbs the molecule upon its first contact with the hyperplane $\{x^{(1)} = 1\}$.

The molecule's position $\vect{X}_t = (X_t^{(1)}, \ldots, X_t^{(D)})$ evolves according to the stochastic differential equation (SDE)
\begin{align} \label{eq:D-dim-SDE}
    \diff \vect{X}_t = \vect{v} \diff t + \sigma \diff \vect{B}_t,
\end{align}
where $\sigma > 0$ is the dimensionless diffusion coefficient, and $\{\vect{B}_t\}_{t \geq 0}$ is a standard $D$-dimensional Wiener process with independent components.

We define the \emph{first arrival time} as
\begin{align} \label{eq:first-hit-time}
    T \eqdef \inf\{t > 0 : X_t^{(1)} = 1\},
\end{align}
and the corresponding \emph{first arrival position} as 
\begin{equation}
    \vect{X}_T = (1, X_T^{(2)}, \ldots, X_T^{(D)}).
\end{equation}

\begin{figure}[!h]
    \centering
    \resizebox{\linewidth}{!}{%
    \begin{tikzpicture}[
        x={(1.4cm,0cm)},    
        y={(0.5cm,0.4cm)},  
        z={(0cm,1cm)},      
        scale=2.0,          
        >=stealth
    ]

    \def\hitY{0.6} 
    \def\hitZ{0.5} 
    
    \fill[gray!10, opacity=0.7] (0,-1,-1) -- (0,1,-1) -- (0,1,1) -- (0,-1,1) -- cycle;
    \draw[gray!40] (0,-1,-1) -- (0,1,-1) -- (0,1,1) -- (0,-1,1) -- cycle;
    
    \node[anchor=north east, xshift=-3pt] at (0,-1,-1) {\footnotesize \textbf{TX Plane} ($x^{(1)}=0$)};
    
    \node[anchor=east] at (0,0,0) {\footnotesize $\mathbf{x}_0$};
    \fill[black] (0,0,0) circle (1.5pt);

    \draw[->, thick, red!80!black] (0.2, 0.1, 0.2) -- ++(0.4, 0.15, 0.1) 
        node[midway, above, sloped] {\footnotesize Drift $\mathbf{v}$};

    \draw[blue!80, thick, decorate, decoration={random steps, segment length=3pt, amplitude=2pt}] 
        (0,0,0) -- (0.5, 0.3, 0.2) -- (1, \hitY, \hitZ);

    \fill[blue!5, opacity=0.6] (1,-1,-1) -- (1,1,-1) -- (1,1,1) -- (1,-1,1) -- cycle;
    \draw[blue!30, dashed] (1,-1,-1) -- (1,1,-1) -- (1,1,1) -- (1,-1,1) -- cycle;

    \node[anchor=north west, xshift=3pt] at (1,-1,-1) {\footnotesize \textbf{RX Plane} ($x^{(1)}=1$)};

    \draw[->, black] (0, -1.5, -1.2) -- (1, -1.5, -1.2) node[midway, below] {\footnotesize Distance $\lambda=1$ (Normalized)};
    
    \draw (0, -1.4, -1.2) -- (0, -1.6, -1.2); 
    \draw (1, -1.4, -1.2) -- (1, -1.6, -1.2); 

    \draw[->, thin, gray] (-0.8, 0, 0) -- (-0.5, 0, 0) node[right] {\scriptsize $x^{(1)}$};
    \draw[->, thin, gray] (-0.8, 0, 0) -- (-0.8, 0.3, 0) node[right] {\scriptsize $x^{(2)}$};
    \draw[->, thin, gray] (-0.8, 0, 0) -- (-0.8, 0, 0.3) node[above] {\scriptsize $x^{(D)}$};

    \coordinate (Hit) at (1, \hitY, \hitZ);
    \fill[red] (Hit) circle (1.5pt);
    
    \draw[dotted, thin, black] (Hit) -- (1, \hitY, 0) node[below, font=\scriptsize] {$X_T^{(2)}$};
    \draw[dotted, thin, black] (Hit) -- (1, 0, \hitZ) node[left, font=\scriptsize] {$X_T^{(D)}$};
    \draw[dotted, thin, black] (Hit) -- (1, 0, 0); 

    \node[right, align=left, font=\footnotesize] at (Hit) {
        \textbf{FAP} $\mathbf{X}_T$\\
        at time $T$ (FAT)
    };

    \end{tikzpicture}%
    }
    \caption{System model illustration. A molecule is released from the transmitter (TX) at $\mathbf{x}_0$ and propagates via Brownian motion with drift $\mathbf{v}$ until it hits the receiver (RX) hyperplane at $x^{(1)}=1$. The first arrival position (FAP) $\mathbf{X}_T$ and first arrival time (FAT) $T$ are recorded upon absorption.}
    \label{fig:system_model}
\end{figure}

\begin{table}[!t]
\caption{Summary of Notation}
\label{tab:notation}
\centering
\begin{tabular}{l l}
\toprule
\textbf{Symbol} & \textbf{Meaning} \\
\midrule
$D$ & Spatial dimension of the environment \\[0.3em]
$t,\, T$ & Time / first arrival time (FAT) \\[0.3em]
$\vect{X}_t$ & Molecule position at time $t$ \\[0.3em]
$\vect{X}_T$ & First arrival position (FAP) \\[0.3em]
$\vect{v} = (1, v^{(2)},\ldots, v^{(D)})$ & Drift velocity vector \\[0.3em]
$\vect{x}^{(2:D)}$ & Lateral coordinates (dimensions $2$ to $D$) \\[0.3em]
$\vect{\Delta}_t$ & Residual lateral displacement at time $t$ \\[0.3em]
$\sigma$ & Diffusion coefficient (dimensionless) \\[0.3em]
$f_{T,\vect{X}_T}$ & Joint FAT--FAP probability density \\[0.3em]
$\mathbf{I}(\vect{\theta})$ & Fisher information matrix (FIM) \\[0.3em]
\bottomrule
\end{tabular}
\end{table}

\section{Joint Statistics and First Arrival Position}
\label{sec:joint_stats}

We analyze the diffusive channel in a $D$-dimensional unbounded
medium with constant drift velocity $\vect{v}$. To derive the joint
statistics of the first arrival time $T$ and first arrival position
$\vect{X}_T$, we decompose the motion into longitudinal and transverse
coordinates:
\[
    \vect{X}_t = \bigl(X_t^{(1)}, \vect{X}_t^{(2:D)}\bigr),
    \qquad
    \vect{x} = \bigl(x^{(1)}, \vect{x}^{(2:D)}\bigr).
\]
The SDE in \eqref{eq:D-dim-SDE} separates componentwise as
\begin{align}
    \diff X_t^{(1)} &= v^{(1)} \diff t + \sigma \diff B_t^{(1)}, \\
    \diff \vect{X}_t^{(2:D)}
    &= \vect{v}^{(2:D)} \diff t + \sigma \diff \vect{B}_t^{(2:D)},
\end{align}
where $B_t^{(1)}$ and $\vect{B}_t^{(2:D)}$ are independent Brownian motions.
Recalling the definition in \eqref{eq:first-hit-time}, the first arrival time $T$ is determined solely by the longitudinal process $X^{(1)}_t$. Thus, $T$ is measurable with respect to $\{B_s^{(1)}\}$ only, and is independent of
the transverse trajectory $\{\vect{X}_t^{(2:D)}\}_{t\ge0}$. This extends
the componentwise independence of Brownian motion
(cf.~\cite[Thm.~2.37]{morters_brownian_2010}) to the first-arrival setting.

Consequently, conditioning on $T=t$ enforces only $X_t^{(1)}=1$, while
the transverse coordinate retains its unconditional Gaussian law. For
any $t>0$ and $\vect{x} = (1,\vect{x}^{(2:D)})$, the joint probability density function (PDF)
factorizes as
\begin{align}
    f_{T,\vect{X}_T}(t,\vect{x})
    &= f_T(t)\, f_{\vect{X}_T \mid T}(\vect{x}\mid t) \nonumber \\
    &= f_T(t)\,
       \delta(x^{(1)} - 1)\,
       f_{\vect{X}_t^{(2:D)}}(\vect{x}^{(2:D)}).
\end{align}
Here, the term $\delta(x^{(1)}-1)$ explicitly enforces the absorbing boundary condition at the receiver plane. The FAT density is the classical inverse Gaussian law:
\begin{equation}
    f_T(t)=\frac{1}{\sqrt{2\pi\sigma^2 t^3}}
    \exp\!\left(-\frac{(1-t)^2}{2\sigma^2 t}\right)
    \Ind\{t>0\}.
\end{equation}
Since the transverse motion follows drift--diffusion without boundary, its PDF is Gaussian. Namely, 
\begin{equation}
\begin{split}
    f_{\vect{X}_t^{(2:D)}}(\vect{z})
    &= (2\pi\sigma^2 t)^{-\frac{D-1}{2}} \\
    &\quad \times \exp\biggl( -\frac{\|\vect{z} - \vect{x}_0^{(2:D)} - \vect{v}^{(2:D)} t\|^2}{2\sigma^2 t} \biggr).
\end{split}
\end{equation}
Combining both distributions, the joint PDF becomes
\begin{align}
    f_{T,\vect{X}_T}(t,\vect{x})
    &= (2\pi\sigma^2)^{-D/2}\,
       t^{-D/2 - 1}\,
       \delta(x^{(1)} - 1)\,
       \Ind\{t>0\} \nonumber\\
    &\quad \times \exp\biggl( -\frac{1}{2\sigma^2 t} \Bigl[ (1-t)^2 \nonumber\\
    &\qquad\quad + \|\vect{x}^{(2:D)} - \vect{x}_0^{(2:D)} - \vect{v}^{(2:D)} t\|^2 \Bigr] \biggr).
    \label{eq:joint-pdf}
\end{align}
To obtain the First Arrival Position (FAP) density, we integrate
\eqref{eq:joint-pdf} over $t>0$, producing an expression of the form
\[
I=\int_0^\infty t^{\nu-1}\exp\!\left(-\beta t - \frac{\gamma}{t}\right)\diff t,
\]
which evaluates to a modified Bessel function
$K_\nu(\cdot)$ (cf.~\cite{gradshteyn_table_2007}), yielding a closed-form
expression for the FAP distribution in arbitrary dimension.

\section{Fisher Information Analysis}
\label{sec:fisher}

We quantify the information content of the joint observation $(T, \vect{X}_T)$ via the Fisher information matrix (FIM) for the parameter vector $\vect{\theta} = (\sigma, v^{(2)}, \ldots, v^{(D)})$. This analysis establishes fundamental limits on parameter estimation accuracy.

The log-likelihood function corresponding to \eqref{eq:joint-pdf} (excluding constant terms) is
\begin{align} \label{eq:loglike}
    \begin{split}
        &\log f_{T, \vect{X}_T}(t, \vect{x}; \vect{\theta})
        = -\tfrac{D}{2}\,\log (2\pi \sigma^2)
        - \left(\tfrac{D}{2} + 1 \right) \log t 
        \\
        &\quad\quad\quad\quad
        - \frac{(1 - t)^2 + \|\vect{x}^{(2:D)} - \vect{x}_0^{(2:D)} - \vect{v}^{(2:D)} t \|^2}{2\sigma^2 t}.
    \end{split}
\end{align}
Define the residual vector at time $t$ as $\vect{\Delta}_t \eqdef \vect{x}^{(2:D)} - \vect{x}_0^{(2:D)} - \vect{v}^{(2:D)} t$, with $k$-th component $\Delta_t^{(k)}$ for $k \in \{2,\ldots,D\}$.

\subsection{Partial Derivatives}

The partial derivatives of the log-likelihood are summarized below:

\begin{itemize}
    \item \textbf{Derivative w.r.t.\ $\sigma$:}
    \begin{equation} \label{eq:loglike-deriv-sigma}
        \frac{\partial}{\partial \sigma} \log f
        = -\frac{D}{\sigma}
          + \frac{(1 - t)^2 + \|\vect{\Delta}_t\|^2}{\sigma^3 t}.
    \end{equation}

    \item \textbf{Derivative w.r.t.\ $v^{(k)}$, $k \in \{2,\ldots,D\}$:}
    \begin{equation} \label{eq:loglike-deriv-vk}
        \frac{\partial}{\partial v^{(k)}} \log f
        = \frac{\Delta_t^{(k)}}{\sigma^2}.
    \end{equation}
\end{itemize}

\subsection{Fisher Information Matrix Entries}

The FIM entries are defined by $[\mathbf{I}(\vect{\theta})]_{i,j} = -\E_{(T,\vect{X}_T)} [ \partial^2 \log f / \partial \theta_i \partial \theta_j ]$.
For the diffusion coefficient $\sigma$, differentiating~\eqref{eq:loglike-deriv-sigma} yields
\begin{align}
    \frac{\partial^2}{\partial \sigma^2} \log f
    = \frac{D}{\sigma^2} - \frac{3}{\sigma^4 t}\Big((1 - t)^2 + \|\vect{\Delta}_t\|^2\Big).
\end{align}
Taking the expectation, we obtain
\begin{align}
    I_{\sigma\sigma} 
    &= \frac{3}{\sigma^4}\,\E\!\left[
    \frac{(1 - T)^2}{T} + \frac{\|\vect{\Delta}_T\|^2}{T}
    \right] - \frac{D}{\sigma^2}.
\end{align}
Using the moments of the inverse Gaussian distribution, $\E_T[(1 - T)^2/T] = \sigma^2$. Furthermore, conditioned on $T$, $\vect{\Delta}_T \sim \mathcal{N}(0, \sigma^2 T I_{D-1})$, implying $\E [\|\vect{\Delta}_T\|^2 / T] = (D-1)\sigma^2$. Combining these results, we obtain
\begin{align}
    I_{\sigma\sigma} = \frac{2D}{\sigma^2}.
\end{align}
For the cross-terms between $\sigma$ and lateral drift $v^{(k)}$, the mixed derivative is proportional to $\Delta_t^{(k)}$. Since $\E[\Delta_T^{(k)}] = 0$, the off-diagonal terms vanish, i.e., $I_{\sigma, v^{(k)}} = 0$.
Finally, for the drift components $v^{(k)}$, the second derivative is $-t/\sigma^2$. Thus we have
\begin{align}
    I_{v^{(k)} v^{(k)}}
    = \frac{1}{\sigma^2}\,\E[T]
    = \frac{1}{\sigma^2}.
\end{align}
The resulting FIM is diagonal, implying that the joint FAT--FAP observation decouples the information regarding diffusion from that of each drift component. Specifically, diffusion information aggregates from both temporal and spatial dimensions ($2D/\sigma^2$), while each lateral drift component contributes $1/\sigma^2$.

\subsection{Estimator Design and Performance Bounds}
To demonstrate the practical utility of the joint statistics, we consider the problem of sensing the environmental flow velocity and diffusivity using an MC receiver. This corresponds to the estimation of $\vect{\theta}$ given $N$ independent observations $\{(T_i, \vect{X}_{T,i})\}_{i=1}^N$.

\subsubsection{Lateral Drift Estimation}
\begin{figure}[!t]
    \centering
    \includegraphics[width=0.7\linewidth]{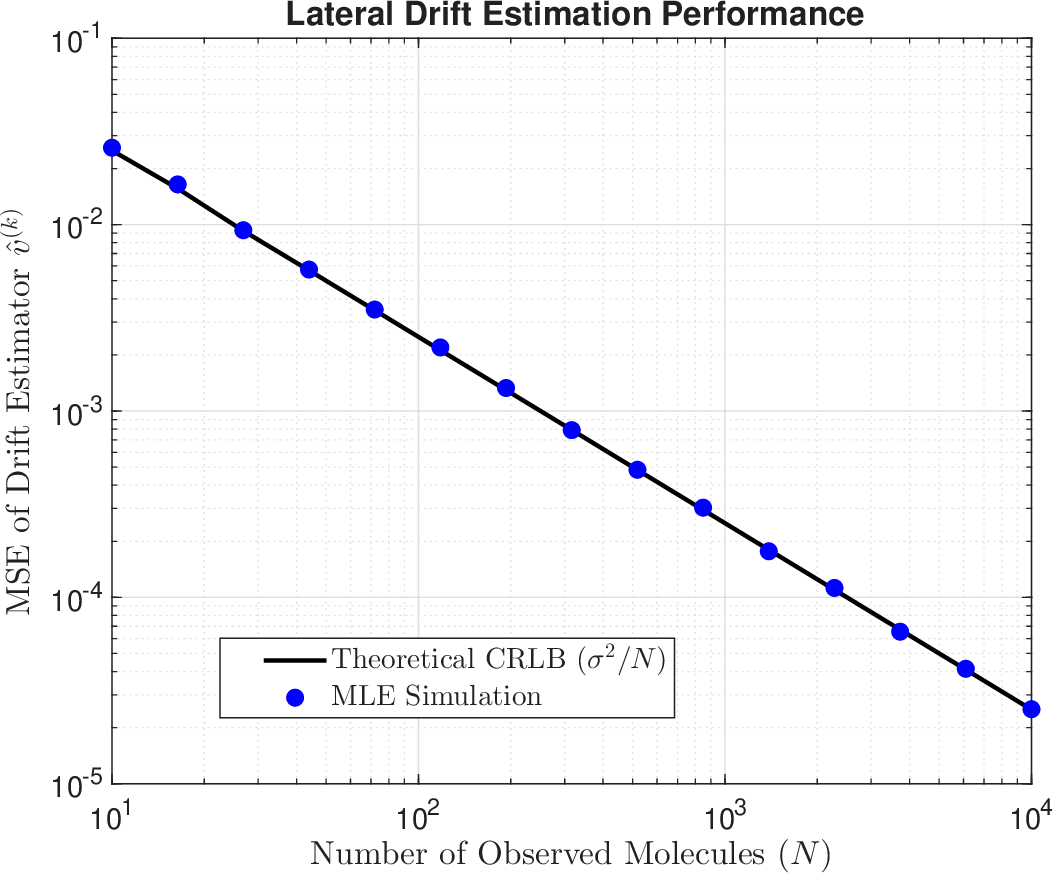}
    \caption{Performance of the derived lateral drift estimator. The simulation results (markers) tightly align with the theoretical CRLB (solid line), validating the efficiency of the proposed MLE in Eq.~\eqref{eq:mle_drift}.}
    \label{fig:drift_mse}
\end{figure}

We derived the MLE for the lateral drift vector $\vect{v}_{\perp}$ by solving $\nabla_{\vect{\theta}} \sum \log f = \vect{0}$, which yields a physically intuitive form:
\begin{equation} \label{eq:mle_drift}
    \hat{\vect{v}}_{\perp}^{\text{MLE}} = \frac{\sum_{i=1}^N (\vect{X}_{T,i}^{(2:D)} - \vect{x}_0^{(2:D)})}{\sum_{i=1}^N T_i}.
\end{equation}
This estimator suggests that the optimal estimate of the background flow is simply the \emph{total lateral displacement} divided by the \emph{cumulative arrival time}. 
Fig.~\ref{fig:drift_mse} compares the Mean Squared Error (MSE) of this estimator against the theoretical CRLB, $\text{Var}(\hat{v}^{(k)}) \geq \sigma^2/N$. The simulation confirms that the proposed MLE is efficient, achieving the fundamental limit even for moderate sample sizes ($N > 100$).

\subsubsection{Diffusivity Estimation and Spatial Advantage}
\begin{figure}[!t]
    \centering
    \includegraphics[width=0.7\linewidth]{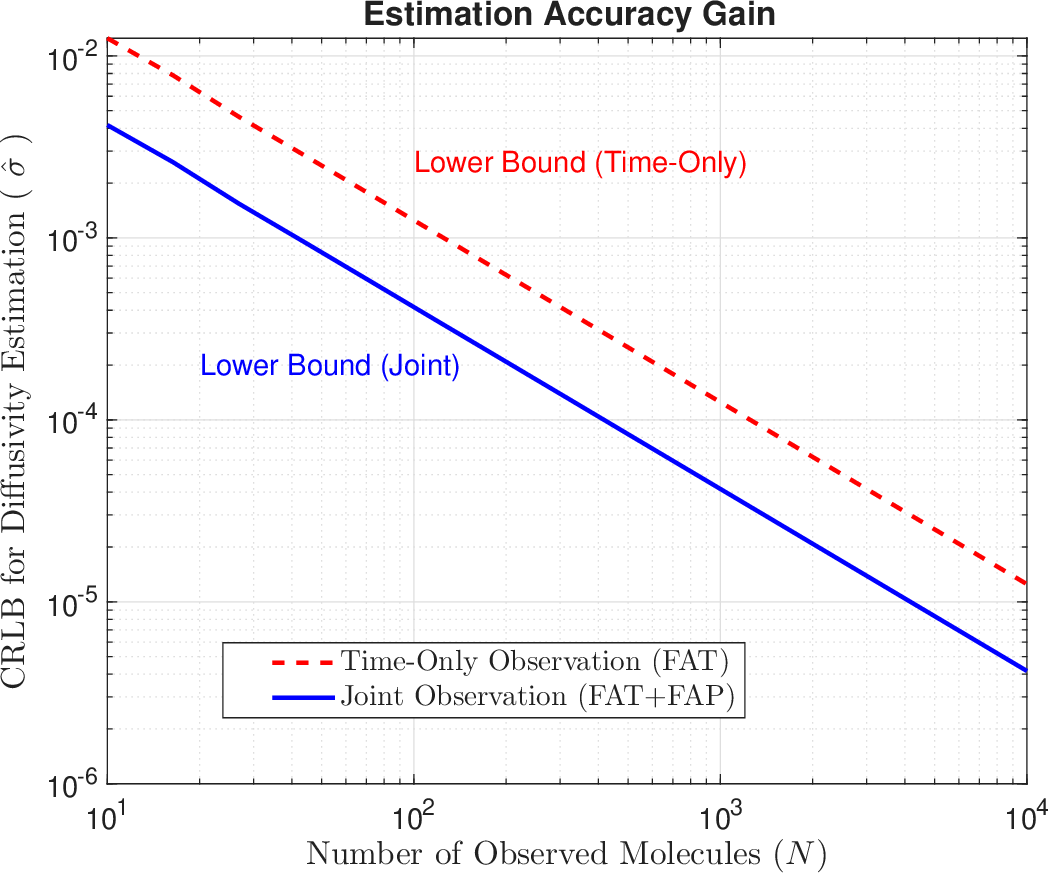} 
    \caption{CRLB comparison for diffusivity estimation ($\sigma$). Joint FAT--FAP observation achieves a strictly lower CRLB than classical FAT-only methods across all sample sizes, indicating that spatial statistics contribute additional Fisher information not present in classical FAT-only models.}
    \label{fig:sigma_mse}
\end{figure}

A key finding of our Fisher information analysis is that spatial observations enhance the sensitivity to diffusivity. The CRLB for $\sigma$ decreases as the spatial dimension $D$ increases: $\text{Var}(\hat{\sigma}) \geq \sigma^2 / (2DN)$.
Fig.~\ref{fig:sigma_mse} illustrates this advantage by comparing the theoretical error bounds of a classical Time-Only receiver ($D=1$, using FAT only) versus the proposed Joint receiver ($D=3$).
The Joint receiver leverages the additional spatial degrees of freedom to significantly lower the estimation error floor. This implies that for a target accuracy (e.g., MSE $\le 10^{-4}$), a joint spatiotemporal sensor requires significantly fewer molecules than a timing-based sensor, thereby improving energy efficiency in nanonetworks.

\subsection{Spatiotemporal Diversity in Signal Detection}
Beyond parameter estimation, the proposed joint statistics enable new degrees of freedom for modulation and detection. Consider a \emph{DSK} scheme where information is encoded in the lateral flow direction. The binary hypotheses are:
\begin{align}
    H_0: \vect{v}_{\perp} = -\vect{u}, \quad H_1: \vect{v}_{\perp} = +\vect{u},
\end{align}
where $\vect{u} \in \mathbb{R}^{D-1}$ is a fixed lateral vector. The longitudinal drift $v^{(1)}=1$ and diffusivity $\sigma$ remain constant.

The optimal detection rule is given by the Log-Likelihood Ratio (LLR) test: $\Lambda(t, \vect{x}) = \log \frac{f(t, \vect{x} | H_1)}{f(t, \vect{x} | H_0)} \gtrless 0$.
Substituting the joint PDF \eqref{eq:joint-pdf}, the terms depending solely on $t$ (including the longitudinal drift and normalization factors) cancel out. The LLR simplifies to:
\begin{align}
    \Lambda(t, \vect{x}) &= -\frac{\|\vect{x}_{\perp} - \vect{u}t\|^2}{2\sigma^2 t} + \frac{\|\vect{x}_{\perp} + \vect{u}t\|^2}{2\sigma^2 t} \nonumber \\
    &= \frac{2 \langle \vect{x}_{\perp}, \vect{u}t \rangle}{2\sigma^2 t} = \frac{\vect{x}_{\perp} \cdot \vect{u}}{\sigma^2}.
\end{align}
The optimal decision rule is thus $\vect{x}_{\perp} \cdot \vect{u} \gtrless 0$.
This result leads to a key system-level insight:
\begin{itemize}
    \item \textbf{Joint Receiver:} The decision boundary is the hyperplane orthogonal to $\vect{u}$ passing through the origin. The error probability is determined by the spatial separation of the Gaussian kernels, providing reliable communication.
    \item \textbf{Time-Only Receiver:} Since the marginal FAT distribution $f_T(t)$ depends only on the longitudinal drift (which is identical under $H_0$ and $H_1$), the LLR for a time-only receiver is theoretically zero. Consequently, a conventional receiver is fundamentally incapable of demodulating lateral drift information ($P_e = 0.5$).
\end{itemize}
This analysis confirms that joint spatiotemporal processing provides a \emph{diversity gain} that transforms previously ``invisible'' channel dimensions into usable communication resources.

\begin{table*}[!t]
\caption{Summary of Fisher Information and System Capabilities: FAT-Only vs.\ Joint Receiver}
\label{tab:performance_summary}
\centering
\begin{tabular}{l c c l}
\toprule
\textbf{Metric / Task} 
& \textbf{FAT Only} 
& \textbf{Joint (FAT+FAP)} 
& \textbf{Insight} \\
\midrule

\rule{0pt}{2.2ex}\textbf{Diffusivity ($\sigma$) FIM} 
& $2/\sigma^{2}$ (eff.\ $D=1$) 
& $2D/\sigma^{2}$ 
& Information scales linearly with dimension $D$. \\

\addlinespace[0.4em]

\rule{0pt}{2.2ex}\textbf{Diffusivity CRLB} 
& $\mathrm{Var}(\hat{\sigma}) \ge \sigma^{2}/(2N)$ 
& $\mathrm{Var}(\hat{\sigma}) \ge \sigma^{2}/(2DN)$ 
& Joint statistics improve variance by factor $D$. \\

\addlinespace[0.4em]

\rule{0pt}{2.2ex}\textbf{Lateral Drift ($v^{(2:D)}$) Est.} 
& Unobservable 
& $\mathrm{Var}(\hat{v}^{(k)}) \ge \sigma^{2}/N$ 
& Enables recovery of the full lateral flow vector. \\

\addlinespace[0.4em]

\rule{0pt}{2.2ex}\textbf{Optimal Drift Estimator} 
& N/A 
& $\displaystyle 
\hat{\vect{v}}^{(2:D)} 
= \frac{\sum_{i} \Delta \vect{x}_{i}}{\sum_{i} t_{i}}
$ 
& MLE has a simple form: displacement / time. \\

\addlinespace[0.4em]

\rule{0pt}{2.2ex}\textbf{Drift Shift Keying (DSK)} 
& Fails ($P_e = 0.5$) 
& Reliable ($\vect{x}^{(2:D)} \cdot \vect{u} \gtrless 0$) 
& Spatial diversity enables modulation invisible to FAT-only. \\

\bottomrule
\end{tabular}
\end{table*}


\section{Numerical Simulations}

We validate the theoretical derivations and evaluate the proposed estimators using numerical simulations, addressing two objectives: validating the joint PDF shape via particle tracking and assessing estimator performance via Monte Carlo trials.

\subsection{Joint PDF Validation}
To verify the closed-form PDF in \eqref{eq:joint-pdf}, we employ Euler--Maruyama simulations ($10^6$ trajectories) with a time step $\Delta t=10^{-3}$~s to ensure negligible discretization error \cite{kadloor2012molecular}. Physical parameters are set to $\lambda=1~\mu\text{m}$ (intracellular scale \cite{farsad2016comprehensive}) and $\sigma=0.5~\mu\text{m}^2/\text{s}$ (viscous environment \cite{nakano2013molecular}), with the transmitter at the origin. Two drift scenarios are considered: $(1,0)~\mu\text{m}/\text{s}$ (longitudinal only) and $(1,-3)~\mu\text{m}/\text{s}$ (with lateral drift).

\subsection{Estimator Performance and Reproducibility}
To evaluate the estimation performance (Section~\ref{sec:fisher}-C), we utilize direct sampling to ensure statistical reproducibility and avoid time-stepping errors. Specifically, arrival times $T_i$ are drawn from the Inverse Gaussian generator (mean $\lambda/v_p$, shape $(\lambda/\sigma)^2$), and conditioned on $T_i$, lateral positions $\vect{X}_{T,i}^{(2:D)}$ are sampled from $\mathcal{N}(\vect{v}_{\perp} T_i, \sigma^2 T_i \mathbf{I})$.
With $D=3, v_p=1,$ and $v^{(2)} = -2~\mu\text{m/s}$, we computed the MSE over $5,000$ independent trials for observed molecules $N$ ranging from $10$ to $10^4$, confirming the MLE's convergence to the theoretical CRLB.

\begin{figure*}[!t]
    \centering
    \includegraphics[width=0.8\linewidth]{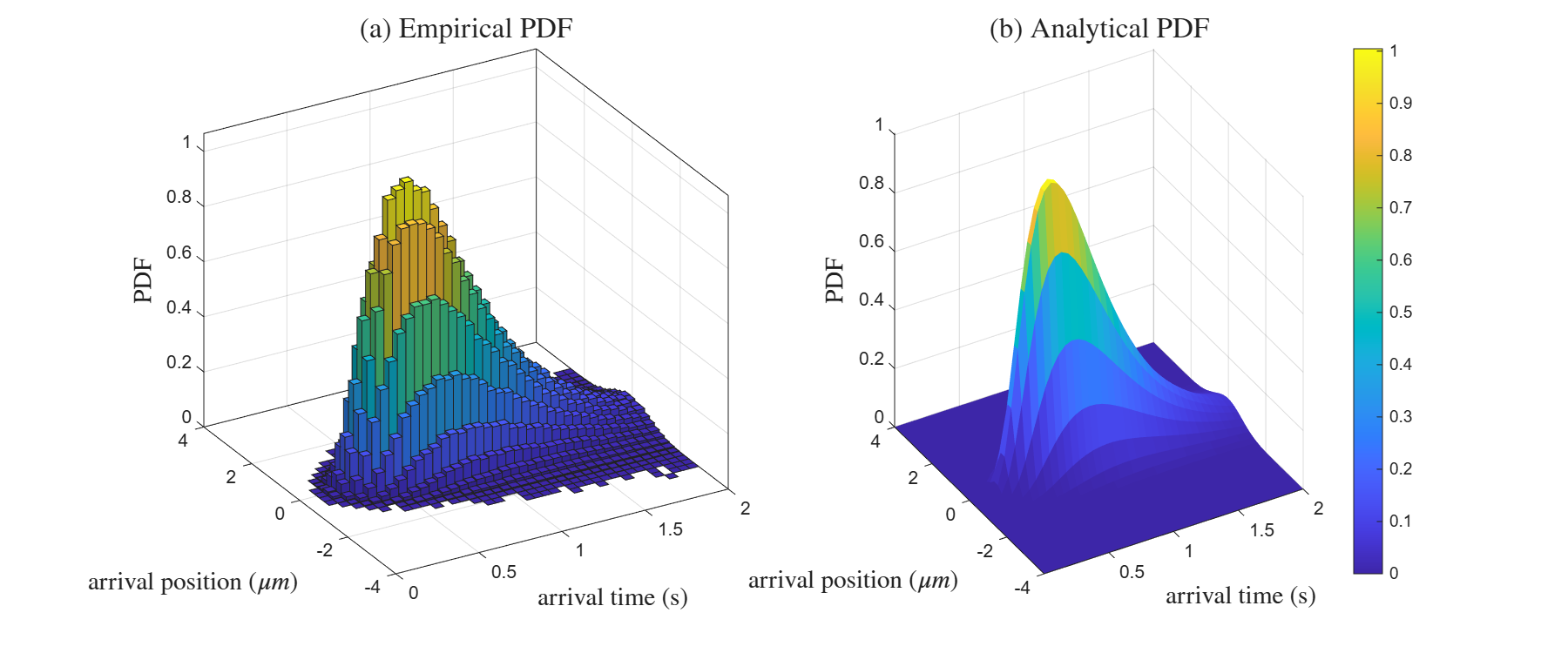}
    \caption{Conditional joint distribution for drift velocity vector $(1,0)~\mu\text{m}/\text{s}$. The figure displays (a) Empirical results from particle-based simulations and (b) Analytical results. The empirical surface accurately reproduces the theoretical peak and decay profile for the longitudinal-only drift scenario.}
    \label{fig:no_drift}
\end{figure*}

\begin{figure*}[!t]
    \centering
    \includegraphics[width=0.8\linewidth]{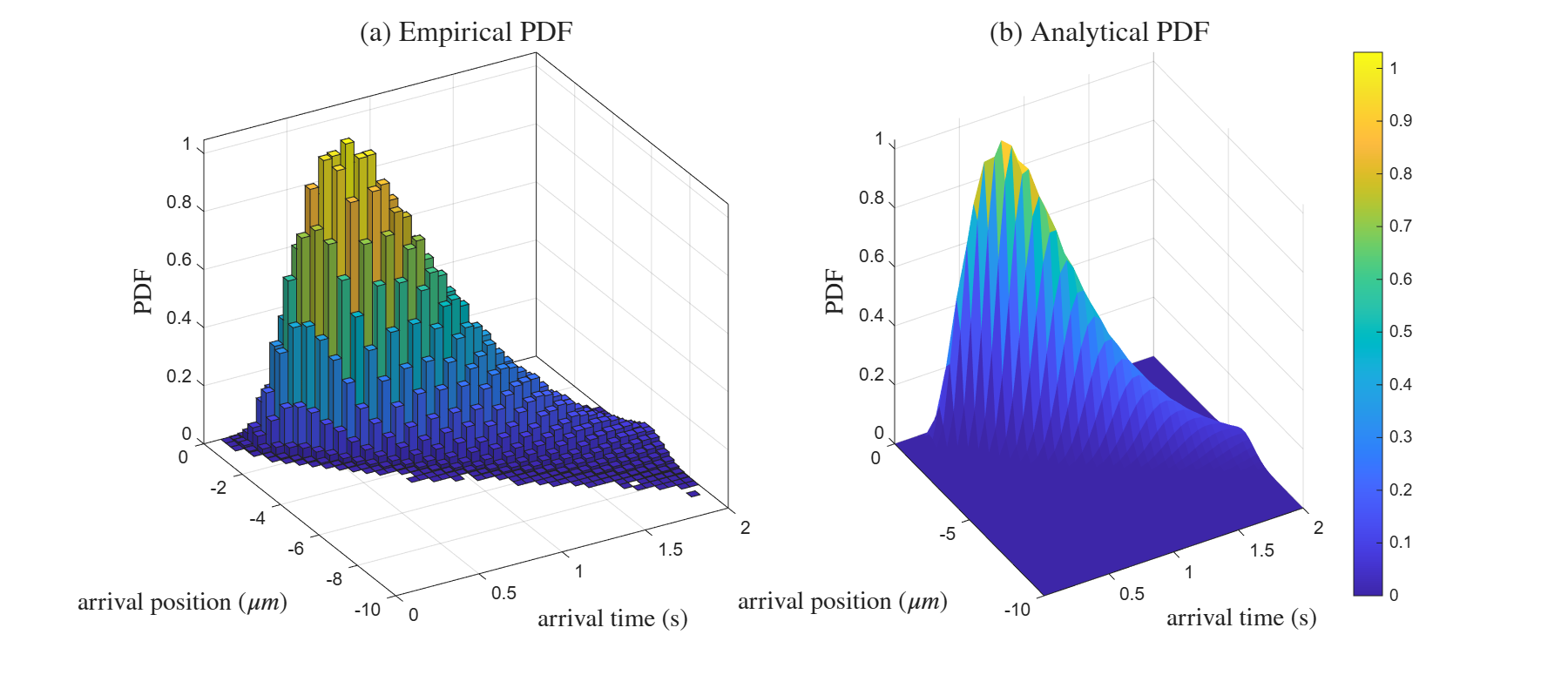}
    \caption{Conditional joint distribution for drift velocity vector $(1,-3)~\mu\text{m}/\text{s}$. (a) Empirical results. (b) Analytical results. The empirical surfaces match the analytical joint PDF not only in peak location but also in the skewed tail induced by the lateral drift component.}
    \label{fig:with_drift}
\end{figure*}

\section{Conclusion}

In this paper, we derived the first closed-form joint distribution for the first arrival time and position in drift-diffusion MC channels, explicitly characterizing the spatiotemporal coupling effects.
Our information-theoretic analysis reveals that spatial observations are not merely supplementary but transformative.
Specifically, we developed a computationally efficient MLE for lateral flow, physically interpretable as the ratio of total lateral displacement to cumulative arrival time.
Furthermore, we demonstrated that exploiting spatial diversity enables novel modulation schemes such as DSK, a regime where traditional time-only receivers are theoretically blind ($P_e=0.5$).
Collectively, these findings suggest that future nanoscale transceivers could benefit from evolving from timing-based architectures to joint spatiotemporal processing designs to fully harness the channel's potential for both communication and sensing.



\balance
\bibliographystyle{IEEEtran}
\bibliography{main}

\end{document}